# Exploring Verbalization and Collaboration during Usability Evaluation with Children in Context

Mohammadi Akheela Khanum[1] and Munesh C. Trivedi [2]

[1] Faculty of Engineering, Department of Computer Science, PAHER University
Udaipur, Rajasthan, India

[2] Department of Computer Science, DIT
Greater Noida, U.P, India

**Abstract**
In this paper, we investigate the effect of context on usability evaluation. The focus is on how children behave and perform when they are tested in different settings. Two most commonly applied usability evaluation methods: the think-aloud and constructive interactions are applied to the children in different physical contexts. We present an experimental design involving 54 children participating in two different configurations of constructive interaction and a traditional think-aloud. The behavior and performance of the children in two different physical contexts is measured by evaluating the results of application of think-aloud and constructive interaction. Finally, we outline lessons on the impact of context on involving children in usability testing.

## 1. Introduction

Now a days when a user buy any gadget, be it a mobile phone, laptop, or an ipad, he first check how easy and understandable the gadget functionality is [1]. This indicates that the users nowadays are more particular about the usability of the gadgets. Usability is most often defined as the ease of use and acceptability of a system for a particular class of users carrying out specific tasks in a specific environment [2]. Ease of use affects the user's performance and their satisfaction, while acceptability affects whether the product is used [2].

With the rapid emergence of new technologies in everyday activities, it is common for all age groups to use new devices. Children cannot be left behind when the use of technologies is discussed. Many children nowadays are found to spend hours with the devices such as laptop computers, game consoles, cell phones, digital cameras, or audio players. All these technologies are becoming an essential part of daily lives. "While many adults struggle with comprehending and manipulating digital interfaces, today's young children enthusiastically approach these interfaces with little or no effort, although they may not completely understand how to use it, or what their implications are" [3].

Children are not miniature adults but they have their own set of preferences, perception, style, likes, and dislikes [4]. When designing technology for children their preferences should be taken into account. To do so, usability evaluation[*] is performed with the children as the testers of technology. During the early design phases of children technology, usability engineers performs usability testing to uncover usability problems that might creep into the product when set to be used in the real context.

Context is a term defined differently by different people. For example, Brown et al. [5] define context as *"location, identities of the people around the user, the time of the day, season, and temperature"*. Ryan et al [6] define context as the *"user's location, environment, identity and time"*. Hull et al [7] included the entire environment by defining context to be *"aspects of the current situation"*. Schilit et al [8] claim that the important aspects of the context are: where you are, who you are with, and what resources are nearby. Dey et al [9] define context to be the *"user's physical, social, emotional or informational state"*.

When evaluating the usability of any system, the behavior of the user is very important. The factors which may affect the user behavior needs to carefully considered because the result of usability evaluations may vary in different settings where the user may exhibit varying behaviors. Product usability doesn't take place in a vacuum; rather, it happens in context [10]. The characteristics of the context (the users, tasks, and environment) may be as important in determining usability as the characteristics of the product itself. Changing any relevant aspect of the context of use may change the usability of the product [11].

Therefore, in this paper we try to find the answer to the following question (i) how does physical context affect verbalizations of perceptions, thoughts, and understandings concerning the interaction in usability evaluations? We address the above stated question by looking at how children perform and behave in lab and in field testing when constructive interaction and think-aloud, methods are applied.

First, we present the literature review on the effect of context during usability testing. Secondly, an experimental design involving 54 children participating in two different configurations of constructive interaction and a traditional think-aloud is presented. Thirdly, we present results from the evaluations by illustrating how the children behaved and perceived the different context when we applied the constructive interaction and think-aloud protocol. Finally, we outline lessons on impact of context on children in usability testing.

## 2. Related Work

The importance of physical context in usability evaluations have been researched for a long. Out of the many factors that can effect usability evaluations, physical context is considered to directly influence the behaviour of the people involved in the usability evaluations. The physical context may include the location, the temperature, the time, the light etc.

Tsiaousis & Giaglis [12] examined the effects of environmental distractions on mobile website usability. They proposed a model hypothesizing on the effects of environmental distractions on the usability of mobile sites. They categorized the environmental distractions into auditory, visual and social. A preliminary test on 20 users was conducted to investigate the effect of environmental distractions on mobile website usability. Results confirmed that environmental distractions have direct effect on mobile website usability.

Hummel et al. [13] developed a mobile context-framework based on a small wireless sensor network, to monitor environmental conditions such as light, acceleration, sound, temperature, and humidity during the usability experiments. User experiments have been conducted in a laboratory with seven test persons where the environmental conditions were changed. Under varying environmental conditions the performance of the users on the average was decreased in terms of higher error rates and delays.

Kaikkonen et al. [14] carried out usability testing of mobile consumer application in two environments: in a laboratory and in a field with a total of 40 test users. Results indicate that conducting a time-consuming field test may not be worthwhile when searching user interface flaws to improve user interaction. They found that field testing is worthwhile when combining usability tests with a field pilot or contextual study where user behavior is investigated in a natural context.

Razak et al. [15] conducted usability testing with children in both laboratory and field. Drawing applications were tested in their preschool and an educational game was tested in the usability laboratory. The results indicate that field study is more suitable for understanding children experience with technology than it is with testing for usability problems and laboratory study is more suitable for evaluating user interfaces and interaction with the application than it is with understanding children's experience.

Andrrzejczak & Liu [16] examined the effect of location on the user's stress level during usability evaluation. User stress levels were assessed by Spielberger's State-Trait Anxiety Inventory; using the paper survey's baseline and experimental stress scores. In addition, user performance data was recorded through task times and subjective user assessments. The data suggested no significant differences exist between participant data in both baseline and experimental anxiety scores. This implies that remote testing as a cost-efficient way to conduct user testing, may be a viable alternative to traditional lab testing without altering the test's effectiveness.

Madathil [17] performed a synchronous remote usability test using a three-dimensional virtual world, and empirically compared it with WebEx, a web-based two-dimensional screen sharing and conferencing tool, and the traditional lab method. The results suggest that virtual lab method is as effective as the traditional lab and WebEx based methods in terms of the time taken by the test participants to complete the tasks and the number of higher severity defects identified. Test participants and facilitators alike experienced lower overall workload in the traditional lab environment than in either of the remote testing environments.

Baillie & Schatz [18] evaluated a multimodal mobile application through a combination of laboratory and field studies. The users were given a set of four action scenarios to be performed. The results were surprising; only one action scenario was completed in the time frame whereas three out of four action scenarios were completed in lesser time. Error rates were higher in lab than in the field. The reason for such performances by the users could be that the users feel more relaxed in the field.

Donker & Markopoulos [19] studies a comparative assessment of three UEMs namely the Concurrent Think Aloud (CTA), interview and questionnaire. Each of these UEMs requires a different level of verbalization for the children that are performing the evaluation. In order to tests these three evaluation methods, 45 children aged 8-14 years were recruited as the test users. The result indicates that children who think aloud during testing uncover more problems than the children who answer specific questions.

However, to elicit verbal comments the children have to be prompted, which can be an indication that children find it difficult to think aloud. Prompting may cause children feel obliged to mention problems to please the experimenter. This could lead to non problems being reported. The result also suggests that girls thinking out loud report more usability problems than boys.

Baauw and Markopoulos [20] conducted a study to compare UEMs. The study involved twenty four children in the age group of 9-11 year, in the usability testing of the computer game- BioMania. The usability evaluation was carried out to test two UEMs namely the TA and post task interview. The results indicate that there was no significance difference between the problems reported by the two genders. The post task interview allows observation data and verbalization data to be obtained on fly without analyzing tapes. Thus, post task interviews can offer practical benefit at the cost of slightly longer sessions. The number of usability problems identified through the two methods was not significant.

Markopoulos and Bekker [21] presented a framework for characterizing comparative studies of usability testing methods with respect to their appropriateness for children. They found that the ability to verbalize problems in interactions depends on: the ability of translating experiences into verbal statements, on their knowledge of the language and on prior experiences in speaking up to adults. They found that compound tasks and abstract tasks formulations could pose problems to children, as their abstract and logical thinking abilities are not yet fully developed and they are not skilled in keeping multiple concepts simultaneously in mind. The results also indicate that think aloud helps generate more problems reports than questionnaires and interviews.

Vermeeren et al., [22] conducted a study on the use of post task interviewing evaluation technique with 6-8 years old children. The results show that children overall were fairly good at answering the questions. The negative side effects of applying the technique on the outcome of the usability test are minor. Further, the study suggests applying such technique to uncover extra data about possible causes for interaction difficulties. Also to limit the questions by only asking detailed questions about those parts of the design that needs extra attention.

## 3. Method

### 3.1 Participants

54 children (24 girls and 30 boys) at the age ranging from 10 years to 13 years old (Mean M=11.63; Standard Deviation SD=0.88) participated as test subjects in the experiment. All the children were 6th and 7th grade pupils from two different English medium schools in the Lucknow city of India. The children did not receive compensation for their involvement in the experiment. The children were assigned as test subjects to one of the four test setups: as individual testers in the lab and in the field for think-aloud sessions, as pairs in lab and field for constructive interaction sessions. Each individual setup had 9 testers (4 girls and 5 boys), and each paired setup had 9 pairs (4 pairs of girls and 5 pairs of boys), Children were randomly assigned to each of the four test setups. Children in pairs were familiar with each other. Table 1 shows the assignment of children to different setups.

Table 1: 54 children assigned as individual testers in think-aloud and as pairs in constructive inetraction

|  | Constructive Interaction | | Think-aloud | |
|---|---|---|---|---|
|  | *Lab* | *Field* | *Lab* | *Field* |
| **Boys** | 5x2 | 5x2 | 5 | 5 |
| **Girls** | 4x2 | 4x2 | 4 | 4 |
| **Total** | 9x2 | 9x2 | 9 | 9 |

### 3.2 Settings

The sessions were held at the school's campus itself, because the school authorities did not permit us to take the children to the place where the usability laboratory was set. We created two labs, one for field testing sessions, and one for laboratory testing sessions. For the field testing, we used the school's computer lab which the students were familiar with and we tried to keep it as it was used by the children. No restrictions were imposed on the people to move in the lab during the test session. This created a perfect field environment for the children. For testing in lab environment, we setup a usability laboratory in one part of the school. The lab environment was different as compared with the field. Lab was located in a quiet place where people not related with the test sessions were not allowed. The lab was only occupied by the test monitors and the test participants at any given time during the test sessions. Fig 1 depicts the usability test session.

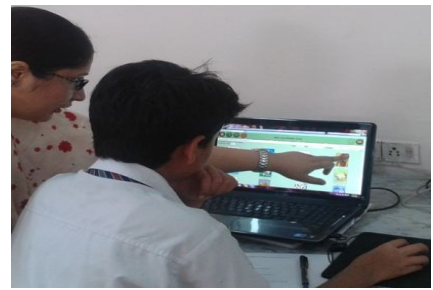

Fig. 1 Snapshot of usability test session

### 3.3 System

The selected system for our experiment was International Children's Digital Library (ICDL). This particular website was selected because digital libraries are becoming a common place for children and many researches are now focusing on how the children are using these new learning tools. During the children's demographic data collection, we also found that none of the children had ever used ICDL. Fig. 2 is the screenshot of ICDL home page.

International Children's Digital Library is a collection of books that features various books for children in different age groups. ICDL has four search tools for accessing the current collection books: Simple, Advanced, Location, and Keyword. Using the simple search, users can search for books using colorful buttons representing the most popular search categories. The advanced search interface allows users to search for books in a compact, text-link-based interface that contains the entire library category hierarchy. By selecting the location based search, users can search for books by spinning a globe to select a continent. Finally, with the Keyword search, users search for books by typing in a keyword.

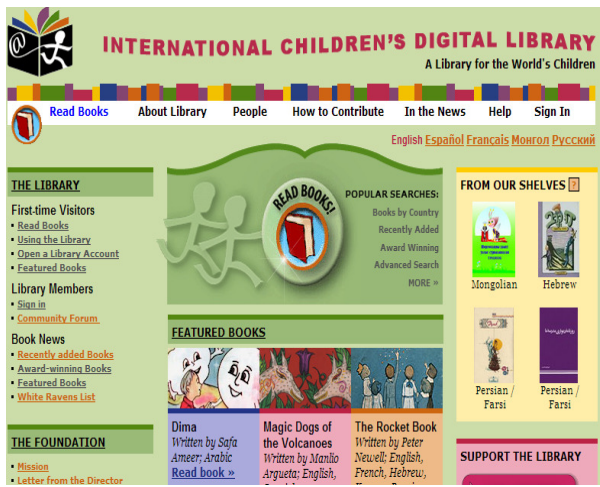

Fig. 2 Screenshot of ICDL

### 3.4 Procedure

The first step towards starting the test was to take consent from the school authorities. After clearing the first step we proceeded with taking the consent from the children's parents or guardians. To do so, we handed over the consent forms to the children to get it signed by their parents or guardians. The consent form provided information about the type of test their wards will be involved in and that the choice of allowing their children to take the test was purely voluntary. After receiving consent from 54 children, we scheduled the usability evaluation sessions. At the beginning of the test sessions children were introduced to the experiment by the participating researchers. The researchers explained the children's roles in the experiment and how their participation would contribute to our research.

Hanna et al. [23] guidelines for usability testing with children were followed. We greeted and children and introduced ourselves. Particularly, we focused on stressing the importance of the participation, and stressing that they were not the object of the test. The purpose of the usability test was explained to the children in detail. The children received questionnaires on which they had to provide answers to such as age, name, school, computer/internet experience, number of hours spend each week on computer/internet, and online reading experience. The usability test sessions were conducted in two labs, one a specialized usability laboratory setup in the school and the other was the school's computer lab. During the test sessions, all the screen activities and children's interaction with ICDL were recorded using CamStudio for later analyses. CamStudio is an open source desktop screen recorder

The children were asked to solve five tasks. The tasks involved the use of different search options in ICDL. This included searching books by country, searching books by title, searching books by language, searching award winning books in English and reading a specified book in the language of their preference. We did not specify any time limits for the tasks, but required the participants to try to solve all tasks.

All children were able to solve all specified tasks. On an average, the children spent 11:11 minutes (SD=2:87) in the lab and 9:33 minutes (SD=2:28) in the field on the all the tasks. The individual testers were asked to think-aloud while solving the tasks.

Think-aloud was explained to the individual testers in terms of the descriptions in [24]. The pairs were asked to collaborate with each other while solving the tasks. Constructive interaction was explained to the pairs as described in [24].

After the usability sessions, the children were asked to complete the subjective workload test (NASA-TLX) [25]. The children filled in the test form individually even though they participated in pairs. NASA-TLX is applied to evaluate the workload as experienced by the children in order to compare their behavior in different settings.

## 4. Data Analysis

36 sessions were completed and then analyzed in detail. The sessions were analyzed based on how well children verbalized (in think-aloud sessions) and collaborated (in constructive interaction sessions). The different aspects of our analysis were (i) Degree of verbalization and collaboration, (ii) Quality of verbalization and collaboration, (iii) impact of test monitor on solving the tasks, (iv) communication between the test monitor and the user and (v) prompting by the test monitor. The quantitative values were assigned to each of these parameters on a scale of 1 to 5. A score of 1 means the lowest and 5 means the highest. For instance, a score of 5 assigned to verbalization/collaboration means that the children verbalized their thoughts to the maximum during think-aloud sessions and collaborated highest during constructive interaction sessions.

## 5. Results

The 54 children in the 36 usability test sessions solved all the assigned tasks. The task completion time in the field (M=9.78, SD=2.28) was lesser compared to the time taken in the laboratory (M=10.67, SD=2.87). But no significance difference was found for the task completion times.

5.1 Assessment of verbalization and collaboration in different settings

To assess the four setups we applied six different aspects of verbalization and collaboration in usability tests. These six aspects are illustrated in table 2. The setting whose mean score (M) marked with a plus sign indicates that it has a significant difference with the setting whose M is marked with a minus sign. SD is the standard deviation. Verbalization refers to the verbal comments during think-aloud sessions which would facilitate identification of what the tester is feeling about the interface under test. Collaboration refers to verbalization during constructive interaction sessions.

Interestingly, we found that the quality of verbalization was considerably higher for the constructive interaction sessions compared to the think-aloud sessions. The score in the lab (M=4.0, SD= 0.5) and in the field (M=3.8, SD=0.4) did not differ much amongst the pairs. However, the score was higher in the field (M=2.67, SD=0.67) as compared to lab (M=1.89, SD=0.74) for the individual testers.

The analysis of variance shows significant differences between the four settings on degree of verbalization $F(3, 32) = 22.55$, $p= 4.93811E-08$. Since the value of p indicated a significant difference between the settings, we performed a post-hoc test.

The post-hoc analysis showed significant difference at the 1% and 5% level between the pairs and individual testers in the lab and the field during both the constructive interaction and think-aloud sessions, however the difference was not significant amongst the pairs and amongst individual testers in the four settings.

Further, we analyzed the quality of verbalization and collaboration in the test sessions. The quality of the collaboration was higher for both the constructive interaction sessions than the quality of verbalization for think-aloud sessions. Field settings provoked more verbalization and collaboration for the testers. The analysis of variance shows significant difference between all the setups on the quality of verbalization/collaboration $F(3, 32) =11.76$, $p=2.35463E-05$. The post hoc analysis showed a significant difference at 1% level between the constructive interaction lab setting and think-aloud lab setting, between constructive interaction field and think-aloud lab setting. At 5% level between constructive interaction lab setting and think-aloud lab setting, between constructive interaction field setting and think-aloud lab setting and also between constructive interaction field and think-aloud field setting.

Table 2: Assessment of verbalization and collaboration in four settings for all testers

| *Testing parameters* | *Constructive Interaction* | | *Think-aloud* | |
|---|---|---|---|---|
| | *Lab* | *Field* | *Lab* | *Field* |
| Degree of verbalization/collaboration | M=4.0+ SD=0.5 | M=3.8+ SD=0.4 | M=1.89- SD=0.74 | M=2.67- SD=0.67 |
| Quality of verbalization/collaboration | M=3.2+ SD=0.8 | M=3.4+ SD=0.5 | M=1.67- SD=0.67 | M=2.44- SD=0.68 |
| Impact of test monitor on solving the tasks | M=2.22 SD=0.67 | M=2.33 SD=0.71 | M=2.56 SD=0.88 | M=2.56 SD=0.53 |
| Communication between test monitor and user | M=2.33 SD=0.50 | M=2.11 SD=0.60 | M=2.44 SD=0.88 | M=2.56 SD=0.53 |
| Prompting by the test monitor | M=2.22+ SD=0.67 | M=2.22+ SD=0.67 | M=3.11- SD=0.33 | M=3.00- SD=0.71 |
| Time taken to complete the tasks | M=10.67 SD=3.67 | M=8.89 SD=2.24 | M=11.56 SD=1.88 | M=9.78 SD=2.17 |

The test monitor plays an important role during usability evaluation. Test monitor is a person who closely monitors the usability test activities and notes the tester's behavior, verbalization, and other such things which may of interest for the usability test under consideration. We analyzed the impact of test monitor on solving the usability tasks. Constructive interaction provides potentially natural thinking-aloud as test subjects collaborate in pairs to solve tasks and therefore, one could expect less influence and interaction with a test monitor. We found that the test monitor has slightly more interaction with the think-aloud subjects compared the constructive interaction subjects, but the difference is not significant $F(3, 32) = 0.5, p=0.684$.

Another factor of our analysis was to assess the level of communication between the test monitor and testers. Test monitor have a slightly higher level of interaction with the testers during think-aloud sessions. However, this difference was not significant $F(3, 32) = 0.78, p=0.515$. We also assessed the level of prompting that was required to make the testers verbalize their actions during the test sessions. Think-aloud required higher level of prompting than the constructive interaction. Also, field testing using think-aloud required lesser prompting compared to lab testing. However, for constructive interaction, prompting in field and lab was not significantly different. The analysis of variance shows significant difference between the setups on the amount of prompting by the test monitor $F(3, 32) = 5.60, p=0.003$. The post hoc analysis showed a significant difference at 5% level between the constructive interaction lab setting and think-aloud lab setting, and between constructive interaction field and think-aloud lab setting.

Finally, we assessed the amount of time spent on solving all the tasks during each test session. Not surprisingly, we found that the testers in think-aloud sessions spent more time on solving the tasks. Field sessions took lesser time compared to their lab counterparts. But this difference is not significant $F(3, 32) = 1.71, p=0.183$.

## 6. Discussion
In this section, we discuss the qualitative results from the study. We have identified a number of interesting outcomes related to usability testing in context with children.

Outcome 1: *usability testing in field provides natural environment for children to freely verbalize their thoughts.*
The children freely verbalize their actions and thoughts in field during constructive interaction and also during think-aloud sessions. Field testing also resulted in better quality of verbalization during both constructive interaction and think-aloud sessions compared to their lab counterparts. Lesser interaction between the test monitor and testers was found in field for both constructive interaction and think-aloud sessions. Time taken to complete all tasks was lesser in field.

Outcome 2: *constructive interaction provides better degree and quality of verbalization compared to think-aloud*
During the constructive interaction sessions the children were more relaxed but during think-aloud sessions they were nervous. Individual testing made the children feel that it was they who were tested and not the interface. One of the individual testers was so nervous that he gave up the test. Higher prompting was required for individual testers. Verbalizing thoughts while solving tasks made the children uneasy. In one case when the monitor asked the tester to verbalize his thoughts, he stopped working and began to think. Working in pairs made the children more comfortable. They discussed much before taking a move while solving the tasks. However, in some cases of constructive interaction the dominating tester ignored the other partner. Lesser intervention by the test monitor was noticed for constructive interaction sessions.

## 7. Conclusion
In this paper, we investigate how children perform and behave in different physical settings during usability testing. Our particular focus is on how the children behave and perceive a testing situation when involved in lab and field testing session with traditional think-aloud and constructive interaction. Our results show that field testing with children resulted in better level and quality of verbalization. Field testing can be a feasible option for testing with children. Even though we did not impose any time constraints on the children, our results show that field testing took lesser time to complete the tasks.

Our results also show that the pairing of children had impact on how the children verbalized and collaborated in pairs during the testing sessions. We found that constructive interaction facilitate natural think-aloud as the pairs tended to collaborate well while solving the tasks. The quality of verbalization was fair enough to get them closer to the solution.

We further experienced that the individual testers applying think-aloud tended to be more verbose in the field than in the lab. This could be an indication that it is not only the method that is affecting the usability tests but also the context in which the test is performed.

Our future goal is to further investigate the impact of context by applying other quantitative measures.